\input{aipcheck}

\documentclass[
    ,final            
  ]
  {aipproc}

\layoutstyle{6x9}

\begin{document}

\title{Exciting Baryons: now and in the future}


\author{M.R. Pennington}{
  address={Theory Center, Jefferson Laboratory, 12000 Jefferson Avenue, Newport News, VA 23609, U.S.A.}
}

\begin{abstract}
 This is the final talk of NSTAR2011 conference. It is not a summary talk, but rather a looking forward to what still needs to be done in excited baryon physics. In particular, we need to hone our tools connecting experimental inputs with QCD. At present we rely on models that often have doubtful connections with the underlying theory, and this needs to be dramatically improved, if we are to reach definitive conclusions about the relevant degrees of freedom of excited baryons. Conclusions that we want to have by NSTAR2021.
\end{abstract}

\maketitle


\section{Where we are}

One cannot look to the future of the baryon physics program without reviewing where we are at present. For the past 50 years we have sought to understand the spectrum of baryons.
It is sometimes thought that spectroscopy is nothing more than stamp collecting, making pretty patterns, but providing few insights into the workings of the world. This is to misunderstand its  fundamental importance.
The spectrum of states of any system are determined by the constituents that make up that system and the forces that 
bind them together. Thus the spectrum provides us with insights into the fundamental degrees of freedom and into the nature of strong coupling QCD.  

Baryons have a special place
in the panoply of hadrons, as their structure is most obviously related to the color degree of freedom.
While a color singlet quark-antiquark system is basically the same however many colors there are, the
minimum number of quarks in a baryon is intimately tied to the number of colors. If $N_c$ were five, the world would 
be quite different. Moreover the flavor pattern of baryons was a key ingredient in the development of the  quark model. This simple model has long served as the paradigm for what we expect the baryon spectrum, both nucleons and $\Delta$'s, to look like~\cite{isgur}.
\begin{figure}[t]
  \includegraphics[height=.4\textheight]{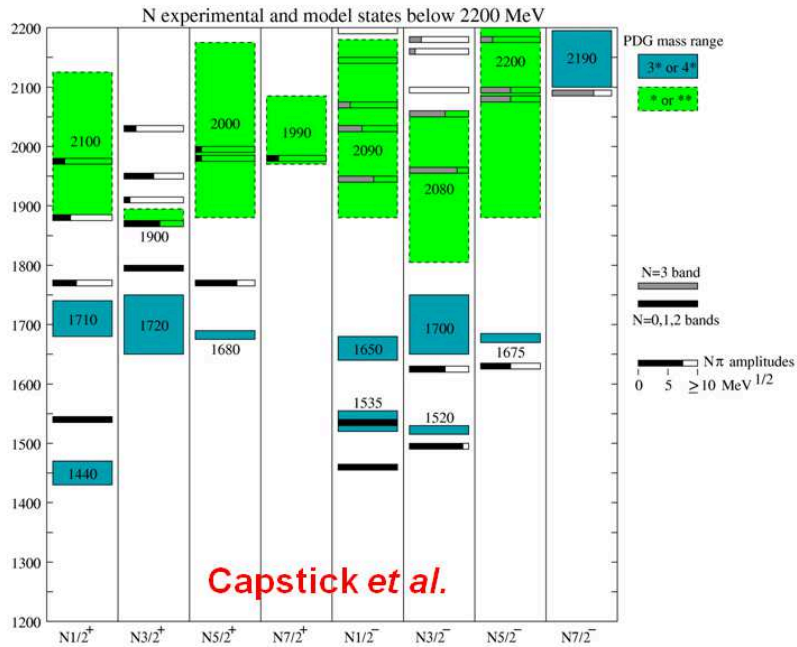}
\vspace{4mm}
  \caption{$N^*$ spectrum, labeled by their spin and parity as $J^P$ along the abscissa from the quark model predictions of Capstick and Roberts~\cite{capstick}. For each state, its $\pi N$ branching fraction is shown. These are compared with the PDG judgement~\cite{pdg} of which states have been identified with $1^*-2^*$ or $3^*-4^*$ provenance, according to the legend shown. }
\end{figure}
The most quoted template, Fig.~1, was provided by Capstick and Roberts~\cite{capstick}, based on three independent quark degrees of freedom, Fig~2.
While the lower lying states have been well determined by experiment, many of those above 1.6 GeV or so are {\it missing}.
A possible reason for this could be that  most of the early evidence was accumulated from $\pi N$ scattering, and decays into the same channel.
Perhaps these states are {\it missing} because we have not looked in the right place. Consequently, there has been a major effort to investigate other channels like $\pi\pi N$ and $KY$. These are an increasing part of the $\pi N$ total cross-section as the energy goes up. However, for the most part they have only contributed hints and glimpses of missing states
and not yet too much new definitive evidence. 

\newpage

\noindent Perhaps these missing states really do not exist.
If baryons were diquark--quark systems, Fig.~2, as Lichtenberg and Tassie noted more than 40 years ago~\cite{lichtenberg}, the number of states  would be restricted and in fact be very like that currently observed. 
\begin{figure}[b]
  \includegraphics[height=.15\textheight]{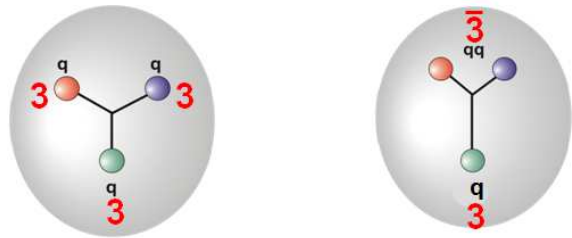}
  \caption{In QCD each quark is in a triplet of color. On the left  is a three quark model of a color singlet baryon. On the right is a diquark-quark model of a baryon, where the diquark must be in a color anti-triplet, so that in this model the baryon is like a meson as far as color is concerned.  }
\end{figure}

But what has this do with QCD?
There are several ways to approach that question. The lattice provides a modeling of the real world with a discretisation of space-time. Whilst the lattice clearly cannot have the complete rotational symmetry of the real world, the newly computed lattice spectrum, as explained by Robert Edwards~\cite{edwards}, reveals a pattern very like
the $SU(6)\times O(3)$ of the quark model: certainly not that of a pointlike diquark--quark system. There is no suggestion that the so far undiscovered states should not be there. However, one essential ingredient presently  missing in such
calculations are explicit continuum states.  Though great computational strides have been made the pion mass is still 3 or 4 times too heavy, and so decay patterns are not yet those of the physical baryons. The inclusion of these coupled channels can have two possible effects: either to increase the richness of the spectrum (examples of which we will discuss later), or conversely states that are there without decays can melt into the continuum once channels are open. Which of these alternatives happens depends on the particular circumstances.

We, of course, know that even a ground state like the $\Delta(1232)$ is not just a three quark baryon. It spends part of its time in a pentaquark configuration, which rearranges itself into a $\pi$ and a nucleon, allowing it to decay. That these are only small components of its wavefunction (or Fock space decomposition) means that the dominant three quark configuration is readily identifiable. However, some $N^*$'s clearly have more equal components of three quark and multi-hadron configurations. Moreover, what role does glue play beyond producing the binding? 

\section{Coupled channels}
A modeling where states and their decays are included from the start as essential ingredients is the program of the Excited Baryon Analysis Center here at JLab~\cite{ebac}. This is an ambitious project that starts from an effective Lagrangian of meson and baryon interactions~\cite{ebac}, and then fits data on $\pi N$ and $\gamma N$ production channels to a host of final states to determine the mass, width and couplings of $N^*$ and $\Delta$ resonances. The EBAC team are on target to complete the analysis of  all possible states below 1.6 GeV or so. This is an heroic effort involving $10^5$ data points. Results already obtained on the $P_{11}$ sector, Fig.~3, show poles in the complex energy plane around 1400 and 1800 MeV, corresponding to the $N^*(1440)$ and $N^*(1710)$ in ~\cite{pdg}. The lower, Roper, state has reflections on two nearby sheets (Fig.~3). Now within the model Lagrangian approach the couplings to decay channels can be tuned. Indeed, if the explicit $\pi\Delta$ and $\eta N$ channels are switched off these $P_{11}$ poles merge into a single entity on the real axis around 1800 MeV~\cite{ebacpoles}.
\begin{figure}[t]
\vspace{-3mm}
  \includegraphics[height=.24\textheight]{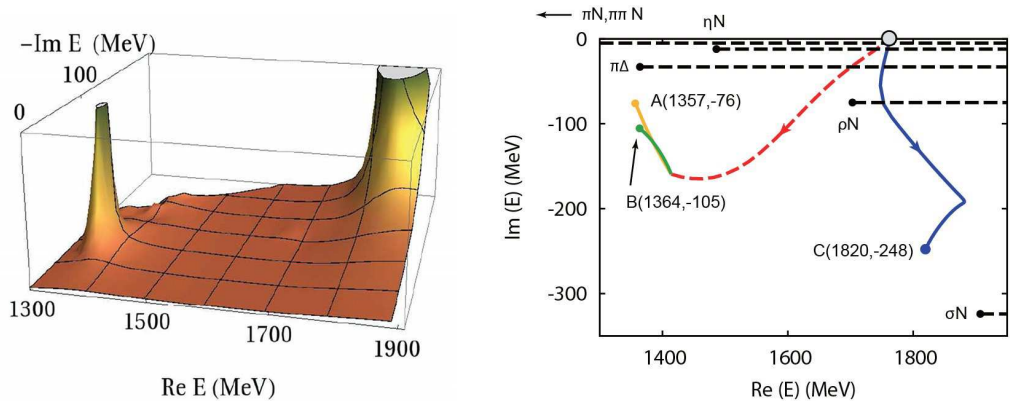}
  \caption{Poles in the $P_{11}$ amplitude in the complex energy plane. On the right the poles are shown in relation to the cuts generated by the channels to which the $N^*$'s decay. The poles A, B are those of the Roper on two different sheets. Pole C is the $N^*(1710)$. When the couplings to the $\pi \Delta$, $\eta N$ are reduced, these poles move from their physical positions along the lines shown towards a single pole on the real energy axis~\cite{ebacpoles}.}
\end{figure}
Within the model this might be regarded as the {\it bare} or {\it seed} pole. 
 This is very interesting, but what does it have to do with QCD? 

A strong coupling model of QCD is provided by the solution of the  Schwinger-Dyson/Bethe-Salpeter (SD/BS) system of equations~\cite{sdbsreviews}. There modeling the gluon by a simple contact interaction (to render the complex of equations tractable) and presuming that baryons are bound states of a quark with an extended (not pointlike) diquark, detailed calculations of the $N^*$ spectrum have been made, as presented here by Craig Roberts~\cite{cdr}. These calculations again have no decays, but amusingly there is a {\it bare} $P_{11}$ state that can be identified with the pole on the real axis of the EBAC zero coupling treatment of Fig.~3 (right). The SD/BS eqution system does allow decay channels to be included, at least in principle, and so future calculations may test whether the evolution from {\it bare} to {\it dressed} states accords with the results of experiment.

That decay channels are critical to the properties and even the very existence of states can be more readily exposed by considering the meson sector. In the world of charmonium, we have long had the notion that a potential model may make sense allowing a combination of short distance one gluon exchange and a longer range confining components. Then for the more tightly bound states below $D{\overline D}$ threshold everything looks under control. However, above that threshold we have learnt that decay channels can play an important role. Indeed, explicit calculation~\cite{wilson,eichten} shows how the states of charmonium are shifted by tens of MeV by their coupling to nearby open or (only just) virtual channels involving $D$'s and $D^*$'s  and their anti-particles. Moreover, the narrow $X(3872)$ may well owe its very existence to forces binding a $D^0$ and a $\overline{D^{*0}}$~\cite{3872}. 

In the light quark sector, a non-relativistic potential picture make little sense. If one looks at mesons with scalar quantum numbers, it has long been known that there are far more states than can fit into a single ${\overline q}q$ multiplet with $S=1$ and $L=1$. Indeed, such a multiplet having a unit of orbital angular momentum more than the ground state vectors, like the $\rho$, should be heavier and more naturally up at 1.3-1.5 GeV. This leaves the lighter scalars, $\sigma$, $\kappa$, $a_0$ and $f_0(980)$ as examples perhaps of tetraquark states, as Jaffe~\cite{Jaffe} proposed already in the '70s. However, let us consider a modeling of states and their decays, as pioneered by van Beveren, Rupp and collaborators~\cite{vanbev}. With a basic ${\overline q}q$ scalar multiplet centred on 1.4 GeV, consider the ${\overline s}s$ state as an example. Switching on decays puts a cut in the propagator function~\cite{pennhadron2009}. This function, no longer has a bare pole on the real axis, but rather two states emerge: one close to 1.4-1.5 GeV with a smallish imaginary part readily identifiable with its ${\overline s}s$ seed, while the other is  pulled down towards $K{\overline K}$ threshold by the strong coupling to this channel, making it naturally identifiable with the $f_0(980)$~\cite{pennhadron2009}. Similar things happen for the other states resulting in the $\kappa$ and $\sigma$ being respectively near to $\pi K$ and $\pi\pi$ thresholds. Such states have almost lost connection to their model ${\overline q}q$ seeds, spending most of their time in di-meson configurations.

 Once again there is an important caveat. These calculations are based on models that are not presently directly related to QCD. Nevertheless, one can more generally ask how do we know if a state is generated by inter-hadron or inter-quark (or constituent forces). This same question was asked by Weinberg about the deuteron~\cite{weinberg}. Is it a six quark bag, or a bound state of two nucleons? And can we tell the difference? Such considerations have been generalized to the situation of the $f_0(980)$ by David Morgan~\cite{morgan}. A state in the spectrum corresponds to a pole in the complex energy plane. It is the pole on the nearest unphysical sheet that has most influence on physics on the real axis, where experiment is performed. There are reflections (mirror poles) on other sheets too, as with the Roper in Fig.~3.  For a state like the $f_0(980)$, whose dynamics depends on more than one channel, and sits close to $K{\overline K}$ threshold, it may have  poles on two unphysical sheets, both of which are equally nearby and so both may determine what experiment measures.  If the $f_0(980)$ is generated by inter-hadron forces, as with a $K{\overline K}$ molecule, then the data should demand just one pole and this would be on what is called sheet II~\cite{mp}. Because of its coupling to $\pi\pi$, this pole is not on the real energy axis as for a pure bound state, but shifted into the $\pi\pi$ unphysical sheet. However, if the forces are shorter range, as with inter-quark binding (or equally inter diquark binding) then the data would demand two poles: one on sheet II and a reflection on the unphysical $K{\overline K}$ sheet, III, too. Precision data in the neighborhood of the $K{\overline K}$ thresholds (charged and neutral) can tell the difference between these two scenarios.

When this methodology, in which analyticity and unitarity are key, was first applied to the $f_0(980)$ by David Morgan and myself~\cite{mp}, we found that the classic CERN-Munich $\pi\pi$
scattering data, on phase-shifts and inelasticities, in 20 MeV bins could never distinguish between the one and two pole options~\cite{mp}. One needed data in much finer binnings.
\begin{figure}[h]
  \includegraphics[height=.38\textheight]{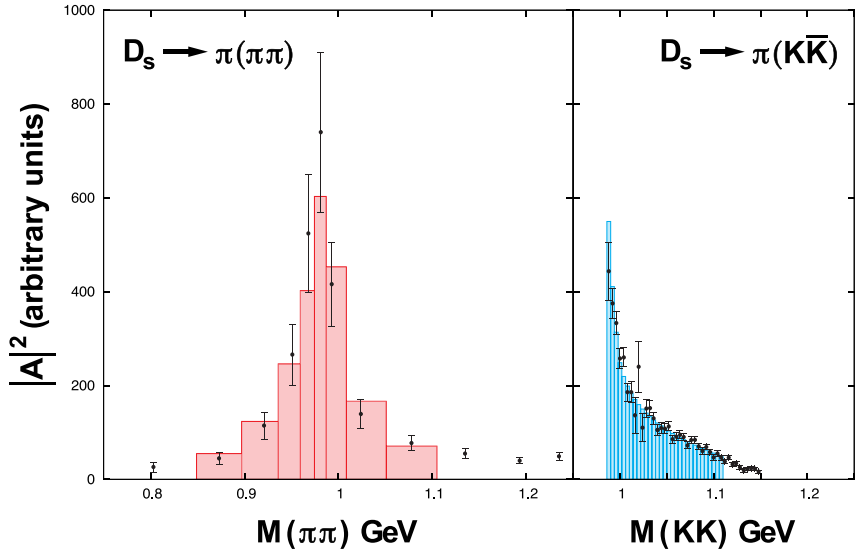}
  \caption{Data-points are the modulus squared of the $S$-wave dimeson amplitudes in $D_s \to \pi(MM)$, where $M=\pi, K$, respectively, from the partial wave analyses of BaBar results~\cite{marco1,marco2}. Note the very fine binning of the $K{\overline K}$ data. The histograms are the result of ~\cite{wilson_f0} using the Jost function methodology of ~\cite{mp}. These favor the single-pole (or molecular) $f_0(980)$. The phase information from the partial wave analyses, which is not shown here is equally critical for the discrimination between 1-pole and 2-pole fitting. The $K{\overline K}$ data are in units of (50 Events per 4 MeV)~\cite{marco2}.}.
\end{figure}
 While both $J/\psi\to \phi(\pi^+\pi^-)$ and $D_s \to \pi(\pi\pi)$ decays show the $f_0(980)$ as a peak, the data   from Mark III, BESI and FOCUS  were still   not precise enough to be discriminants~\cite{mp}. However, now we have the far greater statistics from $B$-factories. This has allowed BaBar to perform a partial wave analysis of $D_s\to \pi(\pi^+\pi^-)$ and $D_s\to\pi(K^+K^-)$ channels, giving both magnitudes and phases~\cite{marco1,marco2} of the dimeson $S$-wave, Fig.~4. That in the $K^+K^-$ channel is in 4 MeV bins. 

David Wilson and I~\cite{wilson_f0} have shown these data strongly favor the one pole option. A second pole anywhere nearby is just not needed.
 This is definite evidence that the $f_0(980)$ behaves as a ${\overline K}K$-molecule, just as Nathan Isgur~\cite{weinstein} had suggested. This very same method could equally be applied to the $\Lambda(1405)$, as Dalitz~\cite{dalitz} had long ago proposed, see also ~\cite{hanhart}. Unfortunately the data on $K^-p\to \pi\pi\Sigma$ and $\pi^- p\to K^0\pi\Sigma$ channels are not precise enough, with fine enough gradation, to be sure only the pole on the  unphysical $\pi\Sigma$ sheet below $KN$ threshold is all that is required. The situation is complicated by the claim from Magas, Oset and Ramos~\cite{oset} that data on the $K^-p$ and $\pi^-p$ channels in fact require two different, but closely spaced, resonances. Whether both  of them are molecule-like is quite unclear. Only data  in few MeV bins can resolve this issue. As seen here for the lightest scalars and quite possibly for the $\Lambda(1405)$, the inclusion of coupled decay channels creates a more complex and richer spectrum than that of the simple quark model. Coupled channels are essential for understanding the spectrum of excited states.  

\section{The future of the excited baryon program}
A way to explore the constituents of matter is through deep inelastic scattering. If we could perform this on every hadronic state, we would readily learn about its degrees of freedom. While at small Bjorken-$x$, every hadron, whether nucleon, pion, glueball or hybrid is the same, at larger $x$ ($> 0.1$) their individuality would be seen. The nearest we can get to such an experiment with unstable targets, is by excitation. 
Hit by a probing photon, like a lightning bolt, a proton can be excited to  $N^*$'s, Fig.~5. By studying such processes as the photon probes deeper can teach us about the properties of each $N^*$~\cite{mokeev,burkert}. To understand what such experiments tell us requires  knowledge of how the internal degrees of freedom of baryons are encoded by QCD. 
\begin{figure}[bh]
  \includegraphics[height=.23\textheight]{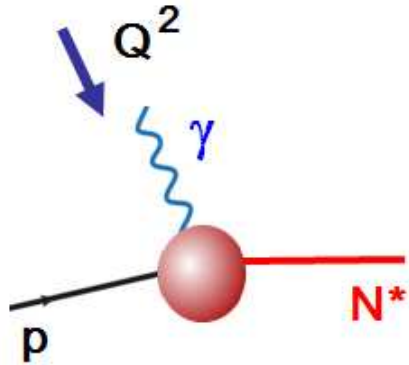}
  \caption{The amplitude for exciting a nucleon into an $N^*$ state using a virtual photon of momentum squared of $-Q^2$. This is produced by inelastic scattering of an electron off the nucleon.}
\end{figure}
As yet, we are still in the modeling stage whether in the quark model, with light cone sum-rules, the Schwinger-Dyson/Bethe-Salpeter approach or on the lattice~\cite{gothe-mokeev}.
  The precision of data emerging from JLab and Mainz makes more focused study of this aspect of strong coupling QCD an imperative for the near future.

As we have stressed $N^*$'s and $\Delta$'s are revealed as much by their decay channels, as their formation reactions.
To determine the role of these channels from experiment requires the appropriate machinery, in which we exploit the fundamental principles of $S$-matrix theory to the full, not just in the narrow energy regions of the $f_0(980)$ or $\Lambda(1405)$, but everywhere. To date too many analyses of data are model-dependent. If their results are ever to be connected to a fundamental theory like QCD, then there are basic properties that these models must fulfill. They must include the consequences of causality, relativistic dynamics and the conservation of probability. These intrisic properties mean that model amplitudes must satisfy analyticity, crossing and unitarity. These properties relate one process to another. One cannot analyse $\pi N\to \pi\pi N$, without being concerned about $\pi N$ and $KY$ final states as well as $\pi\pi$ interactions. Unitarity relates amplitudes for different processes at common energies, either common total centre-of-mass energies, or at common sub-energies of final state interactions. This is to be supplemented by crossing and analyticity, as expressed through dispersion relations, which connect closely related reactions at different energies, in particular high  to low energy. 

The machinery of such relationships was documented 50 years ago. However, it is often only now that data are of sufficient precision, covering sufficient inelastic channels, that the imposition of such properties has become essential. Increasingly photoproduction is being used, both at Jefferson Lab and at Mainz, as a trigger for a wider range of hadronic final states. Amplitudes to describe these processes must be built to satisfy gauge invariance, or at the very least current conservation. As a simple example a process like $\phi\to\gamma\pi\pi$, has been studied by $e^+e^-$ colliders at Novosibirsk and Frascati to probe the nature of the $f_0(980)$. Vanishing phase space means the event distribution must go to zero when the photon energy decreases. That is simple kinematics. However, as emphasised by Achasov~\cite{achasov}, gauge invariance requires the distribution should vanish with the cube of the photon energy and not just linearly. Erroneous conclusions can be drawn if this is not implemented. Many other examples abound. Even with such machinery we cannot be sure that we have unambiguously identified the correct underlying $S$-matrix elements that uniquely  contain the details of the spectrum we want to uncover, particularly for  small partial wave amplitudes. Consequently the program in the baryon sector for {\it complete}, or even {\it over-complete} experiments is essential~\cite{lothar}. There for instance in $\gamma N \to K\Lambda$ we can with polarized beams and the development of appropriately polarized targets (like FROST and HDice), together with the self-analysing power of hyperon decays, measure all the possible observables, and explore whether we can really fully determine all the underlying partial wave amplitudes, and so at least in a limited energy range map out the complete baryon spectrum~\cite{sandorfi}. These energy domains provide anchors for the full $S$-matrix treatment, in which there will inevitably be channels we have not measured and have to model.

Analyses like that of EBAC~\cite{ebac,ebacpoles}, based on the GWU $\pi N$ partial waves of SAID~\cite{arndt}, and its other competitors, like MAID~\cite{tiator} and Bonn-Gatchina~\cite{sarantsev}, provide an opportunity to reveal connections of important coupled channels, as we have discussed with regard to Fig.~3.  As mentioned earlier we can switch channels on and off, but if this is really to increase our understanding, we need to learn what this has to do with the underlyng theory. QCD has only one coupling. The quark-gluon and multi-gluon interactions that drive dynamics are all related. One cannot tune one without the other. Turning off the strong coupling loses all the key elements of the world we want to study: dynamical chiral symmetry breaking and confinement. Is there a deeper connection between QCD and the models used for the analysis of data beyond both embodying basic $S$-matrix principles? This we need to understand urgently. Here the Schwinger-Dyson/bound state equation approach may be most helpful. What is the meaning of tuning out decay channels, removing thresholds, etc.? The continuum approach to strong coupling QCD has the flexibility to teach us what this means. 

There has to be an end to the $N^*$ program. It cannot continue in its present state of uncertainty, of baryons missing or missed. This equally applies to the search for exotic mesons, whether hybrids, glueballs or multiquark states. These too are an essential  part of the JLab 12 GeV program primarily for GlueX, but also for CLAS12, as well as the current COMPASS@CERN.  With the precision of data taken already,  with much more to come, we have to have the right analysis tools to do justice to the wealth of experimental information. That is a priority, requiring theorists and experimentalists from across the globe to collaborate.
The meson and baryon spectra are twin {\it peaks} exposing the workings of strong coupling QCD and confinement physics.  The presently {\it dark} baryons cannot remain in the dark beyond the JLab 12 GeV era. We have a 10-15 year time window in which to definitively deduce  what is the light hadron spectrum from experiment, and  what it is from QCD, and  reach an understanding of how they are truly connected .... or decide it's an intractable problem.


\begin{theacknowledgments}
  It is pleasure to thank my fellow organizers for suggesting I give this talk.
The work was authored in part by Jefferson Science Associates, LLC under U.S. DOE Contract No. DE-AC05-06OR23177. 
\end{theacknowledgments}



\bibliographystyle{aipproc}   


\end{document}